\magnification=\magstep1
\openup 1\jot

\font\list=cmcsc10


\newcount\refno
\refno=0
\def\nref#1\par{\advance\refno by1\item{[\the\refno]~}#1}

\def\book#1[[#2]]{{\it#1\/} (#2).}

\def\annph#1 #2 #3.{{\it Ann.\ Phys.\ (N.\thinspace Y.) \bf#1}, #2 (#3).}
\def\apj#1 #2 #3.{{\it Ap.\ J.\ \bf#1}, #2 (#3).}
\def\cmp#1 #2 #3.{{\it Commun.\ Math.\ Phys.\ \bf#1}, #2 (#3).}
\def\cqg#1 #2 #3.{{\it Class.\ Quan.\ Grav.\ \bf#1}, #2 (#3).}
\def\grg#1 #2 #3.{{\it Gen.\ Rel.\ Grav.\ \bf#1}, #2 (#3).}
\def\jmp#1 #2 #3.{{\it J.\ Math.\ Phys.\ \bf#1}, #2 (#3).}
\def\mpla#1 #2 #3.{{\it Mod.\ Phys.\ Lett.\ \rm A\bf#1}, #2 (#3).}
\def\ncim#1 #2 #3.{{\it Nuovo Cim.\ \bf#1\/} #2 (#3).}
\def\npb#1 #2 #3.{{\it Nucl.\ Phys.\ \rm B\bf#1}, #2 (#3).}
\def\plb#1 #2 #3.{{\it Phys.\ Lett.\ \rm B\bf#1}, #2 (#3).}
\def\pla#1 #2 #3.{{\it Phys.\ Lett.\ \rm A\bf#1}, #2 (#3).}
\def\prb#1 #2 #3.{{\it Phys.\ Rev.\ \rm B\bf#1}, #2 (#3).}
\def\prd#1 #2 #3.{{\it Phys.\ Rev.\ \rm D\bf#1}, #2 (#3).}
\def\prl#1 #2 #3.{{\it Phys.\ Rev.\ Lett.\ \bf#1}, #2 (#3).}
\def\sovj#1 #2 #3.{{\it Sov.\ J.\ Nucl.\ Phys.\ \bf#1}, #2 (#3).}

\overfullrule=0pt       

\def\real{I\negthinspace R}
\def\zed{Z\hskip -3mm Z }
\def\half{\textstyle{1\over2}}

\def\spose#1{\hbox to 0pt{#1\hss}}
\def\lta{\mathrel{\spose{\lower 3pt\hbox{$\mathchar"218$}}
     \raise 2.0pt\hbox{$\mathchar"13C$}}}

\def\N{Nielsen}
\def\O{Olesen}

\hbox{ }
\rightline{gr-qc/9505039}
\rightline {DTP/95/21}
\vskip 0.8truecm

\centerline{\bf ABELIAN HIGGS HAIR FOR BLACK HOLES}
\vskip 0.8truecm

\centerline{ Ana Ach\'ucarro\footnote{$^\spadesuit$}{\sl
Email: anaachu@th.rug.nl }$^{1,2,3}$  
Ruth Gregory\footnote{$^\heartsuit$}{\sl Email:
rg10012@amtp.cam.ac.uk}$^{4,5}$
Konrad Kuijken\footnote{$^\diamondsuit$}{\sl Email:
kuijken@astro.rug.nl}$^{1,6}$}

\vskip 1.0cm
\centerline{\list abstract}
\vskip 3mm

{ \leftskip 5truemm \rightskip 5truemm

\openup -1\jot

 We find evidence for the existence of solutions of
the Einstein and Abelian Higgs field equations describing a
black hole pierced by a Nielsen-Olesen vortex. This
situation falls outside the scope of the usual no-hair
arguments due to the non-trivial topology of the vortex
configuration and the special properties of its energy-momentum
 tensor. By a combination of numerical and perturbative techniques we
conclude that the black hole horizon has no difficulty in supporting 
 the long range fields of  the Nielsen Olesen string.
Moreover, the effect of the vortex  can in
principle be measured from infinity, thus justifying its
characterization as black hole ``hair".

\openup 1\jot

\vskip 1 truecm\it PACS numbers: 04.40.-b, 04.70.-s, 11.27+d, 98.80Cq }

\vskip 15mm

{
\openup -1\jot

\noindent{\it 1. Departamento de F\'\i sica Te\'orica,
Universidad del Pa\'\i s Vasco, Lejona, Vizcaya, Spain }

\noindent{\it 2. Department of Mathematics, Tufts University, Medford,
Massachussetts, U.S.}

\noindent{\it 3. Department of Theoretical Physics, University of Groningen, The
Netherlands}

\noindent{\it 4. D.A.M.T.P. University of Cambridge, Silver St, Cambridge, CB3
9EW, U.K.} 

\noindent{\it 5. Centre for Particle Theory, Durham University, South Road,
Durham, DH1 3LE, U.K.}

\noindent{\it 6. Kapteyn Instituut, P.O.Box 800, 9700 AV, Groningen,
Netherlands}

\openup 1\jot}

\vfill\eject
\footline={\hss\tenrm\folio\hss}

Some years ago a collection of results were proved which established that the
only long range information that a black hole could carry was its
electromagnetic charge, mass and angular momentum. Thus, for example,
lepton or baryon number were not good quantum numbers for black holes,
despite being defined for a neutron star. This set of results came to be
know as the ``no-hair'' theorems and, although not justified, 
gave rise to the much
stronger, but for a while popularly held belief that the
only non-trivial field configurations an event horizon could
support were its massless spin one and spin two charges,
Q,M, and J. Such a picture was not only misleading, but
wrong; in spite of the  structure of
the original ``no-hair'' proofs, the no hair theorems only
claim qualified uniqueness of static or stationary black
hole spacetimes  (see [1] for a discussion of no-hair
folklore).

The extrapolation of the no-hair folklore to include matter fields on the event
horizon has been known to be false for some time, as has the extrapolation to
include quantum effects. Black holes can be
coloured[2-5], i.e.~can support long range Yang-Mills hair. Such solutions
are unstable[6,7], thereby evading the usual no-hair uniqueness theorems,
but they do nonetheless exist. Black holes also carry quantum
hair[8,9], which, although not locally observable, can be inferred via
 an Aharonov-Bohm interference in cosmic strings scattered on
either side of the hole. Of course, the no-hair theorems are classical, but one
might expect some modified argument to apply  to
quantum fields, so how is this long range hair mediated?  More importantly,
also at the semi-classical level, the hair has an effect on the thermodynamics
of the black hole[10] caused by the phase shifts of virtual cosmic string loops
dressing the euclidean horizon of the black hole. The existence of these
euclidean vortices[11] seemed at first sight to be incompatible with some of the
principles of the ``no-hair'' theorems, however, an examination of the
appropriate theorem for the abelian Higgs model[12] revealed some assumptions
which were not satisfied for the euclidean vortices[13]. Finally, it was shown
that a small enough magnetically charged black hole would be unstable to the
nucleation of an SU(2) t'Hooft Polyakov field configuration outside the
horizon[14]. Technically of course, this is not new ``hair'' since the magnetic
charge was already measurable at infinity, however, it illustrates nicely the
difference between the ``hair'' and the short range massive fields which in this
case can live outside and on the black hole horizon.

It seems therefore, that one must be quite specific about what one means by
{\it hair} on black holes, and we shall take the definition to mean a charge or
property of the black hole measurable at infinity; whether or not fields can
live on the horizon we shall refer to as {\it dressing}. The
impression that a black hole horizon can support only
a small number of long range massless gauge fields is somewhat misleading, an
examination of the examples cited above indicates a common 
theme: when non-trivial topology
is present in the field theory, the
situation is more subtle and dressing becomes a real
possibility. In this paper we will examine the case of the
abelian Higgs model, and show that the black hole can
indeed sport long hair, namely a U(1) vortex.

This paper was largely motivated by the work of Aryal, Ford and
Vilenkin (AFV)[15], who wrote down a `solution' for a cosmic string threading a
black hole. More precisely, they wrote down an axisymmetric metric which
consisted of a conical singularity centered on a Schwarzschild black hole.
$$
ds^2 = \left ( 1- {2M\over r} \right ) dt ^2 - \left ( 1- {2M\over r} \right )
^{-1} dr^2 - r^2 d\theta^2 - r^2 (1-4G\mu)^2 \sin^2\theta d\phi^2
\eqno (1.1)
$$

While such a metric was extremely suggestive that a true vortex would thread a
black hole, it addressed only the gravitational aspect of the problem. In view
of the delicacy of dressing the event horizon it seems necessary to analyze
this part of the problem as well. The main potential obstruction to putting a
vortex through a black hole is in having the vortex pierce the event horizon.
Recall that the event horizon is generated by a congruence of null geodesics.
If there exists a static vortex-black hole solution, then this congruence must
remain convergence and shear free throughout the core of the vortex as it
touches the black hole. This in turn translates to a
relation on the stress-energy tensor, namely that $T^0_0 - T^r_r=0$. This is
certainly true at the center of the string, where the energy and tension
balance, but it is not clear that as we move away from the center and the null
vector no longer aligns itself with the string worldsheet, that this balance
will still be maintained.  If a real vortex cannot puncture the event horizon
of a black hole it raises questions as to how black holes and cosmic strings
interact. Do they avoid each other entirely? Or does a black hole `swallow up'
a cosmic string caught in its gravitational grasp? And if a static equilibrium
solution does exist, somehow avoiding the geodesic problem, how does the
vortex pierce the horizon, and how does it circumvent the original no hair
theorems for the Abelian Higgs model?

In this paper we establish that the AFV solution can indeed be viewed as a thin
string limit of some physical vortex by demonstrating that an Abelian Higgs
vortex can thread the black hole. The layout of the paper is as follows: We
first review the self-gravitating U(1) vortex in the next section. In section
three, we examine the question of existence of the vortex in the black hole
background, in the absence of gravitational back reaction. We find an analytic
approximation to the solution for strings thin compared to the black hole, and
present analytic and numerical results for strings of varying widths and
winding numbers. We also examine global strings - a scenario for which we
would not expect a gravitational solution - finding the reaction of the vortex
to the event horizon to be different to that of the local string. In section
four we consider the gravitational back reaction of a single thin vortex,
using the analytic approximation developed in section three. We derive the AFV
metric, but find a subtle difference to their work involving a renormalization
of the Schwarzschild mass parameter - all physically measurable results however
agree. We also comment on the thermodynamics of the string-black hole system.
Finally in section five we summarize and discuss our results, including an
examination of more exotic systems such as a string terminating on a black
hole - which has a gravitational counterpart. We also comment on the dynamical
process of black hole  cosmic string interaction and the second law of
thermodynamics. Finally, we try to answer the question of whether the Abelian
Higgs vortex is true {\it hair}, or whether it is just horizon {\it dressing},
and would be more appropriately dubbed a `wig'.
 
\vskip 2mm

\noindent{\bf 2. The Abelian Higgs Vortex.}

We start by briefly reviewing the U(1) vortex in order to establish notation
and conventions. We take the abelian Higgs lagrangian
$$
{\cal L}[\Phi ,A_{\mu}] = D_{\mu}\Phi ^{\dagger}D^{\mu}\Phi -
{1\over 4}{\tilde F}_{\mu \nu}{\tilde F}^{\mu \nu} - {{\lambda  }\over 4 }
(\Phi ^{\dagger} \Phi - \eta ^2)^2
\eqno (2.1)
$$
where $\Phi$ is a complex scalar field, $D_{\mu} = \nabla _{\mu} + ieA_{\mu}$
is the usual gauge covariant derivative, and
${\tilde F}_{\mu \nu}$ the field strength associated with $A_{\mu}$. 
We use units in which $\hbar=c=1$ and a mostly minus signature.
For cosmic
strings associated with galaxy formation $\eta \sim 10^{15}$GeV and $\lambda
\sim 10^{-12}$.

We shall choose to express the field content in a slightly
different manner, and one in which the physical degrees of freedom
are made more manifest.
We define the (real) fields $X, \; \chi $ and $P_{\mu}$ by
$$
\eqalignno{
\Phi (x^{\alpha}) &= \eta  X (x^{\alpha}) e^{i\chi(x^{\alpha}) } &(2.2a) \cr
A_{\mu} (x^{\alpha}) &= {1\over e}
\bigl [ P_{\mu} (x^{\alpha}) - \nabla _{\mu} \chi (x^{\alpha}) \bigr ]\; .
&(2.2b) \cr
}
$$
In terms of these new variables, the lagrangian
and equations of motion become
$$ 
{\cal L} = \eta ^2 \nabla _{\mu}X \nabla ^{\mu}X
+ \eta^2X^2P_{\mu}P^{\mu} - {1\over{4e^2}} F_{\mu \nu} F^{\mu \nu}
-{ {\lambda \eta ^4}\over 4} (X^2 -1)^2 
\eqno(2.3)
$$
$$
\eqalignno{
\nabla _{\mu}\nabla ^{\mu} X
-  P_{\mu}P^{\mu}X + {\lambda   \eta  ^2\over 2} X(X^2 -1) &= 0
&(2.4a) \cr
\nabla _{\mu}F^{\mu \nu} + 2 e^2 \eta^2 X^2 P^{\nu} &=0\; .
&(2.4b) \cr }
$$
Thus $P_{\mu}$ is the massive vector field in the broken symmetry
phase, $F_{\mu\nu} = \nabla_{\mu} P_\nu - \nabla_{\nu}
P_\mu $ its field strength, and $X$ the
residual real scalar field with which it interacts. $\chi$
is not in itself a physical quantity, however, it can
contain physical information if it is non-single valued, in
other words, if a vortex is present. In this case if the
line integral of $d\chi$ around a closed loop (along which
$X=1$) is non-vanishing, single-valuedness of $\Phi$ then
implies $$ \oint \nabla_\mu \chi dx^\mu  = [\; \chi \; ] = 
{{2\pi}} n \; \; {\rm for \; some\; } n \in \; {\rm \zed}\;
.  \eqno(2.5)  $$
Continuity then demands (in the absence of non-trivial spatial
topology) that $X=0$ at some point on any surface spanning the loop - this is the
locus of the vortex. Thus the true physical content of this model is contained
in the fields $P_{\mu}$ and $X$  plus boundary conditions on $P_{\mu}$  and $X$
representing vortices.

The simplest vortex solution is the Nielsen-Olesen vortex[16], an infinite,
straight static solution with cylindrical symmetry.
In this case, we can choose a gauge in which
$$
\Phi = \eta X_0(R)e^{i\phi} \;\;\; ; \;\;\;
P_{\mu} = P_0(R)\nabla_{\mu}\phi 
\eqno (2.6)
$$
where $R=\sqrt{\lambda}\eta r_{_c}$ in cylindrical polar coordinates $(r_c,
\phi)$.
 The equations for $X$ and $P_{\mu}$ greatly simplify to
$$
\eqalignno{
-X_0  '' - {{X_0  '}\over R} + {{P_0 ^2 X_0 }\over {R^2}}
+ {\textstyle{1\over 2}} X_0  (X_0 ^2 -1) &= 0 &(2.7a) \cr
-P_0  '' + {{P_0  '}\over R} + {{\beta^{-1}}} X_0 ^2 P_0 &=
0 &(2.7b) \cr }
$$
where ${\beta} = \lambda/2e^2 = m^2_{scalar}/m^2_{vector}$
 is the Bogomolnyi
parameter[17] (${\beta}=1$ corresponds to the vortex
being supersymmetrizable). Note that in these rescaled
coordinates, the string has width of order unity.  This
string has winding number one; for winding number $N$, we
replace $\chi$ by $N\chi$, and hence $P$ by $NP$.

This solution can be readily extended to include self-gravity by using Thorne's
cylindrically symmetric coordinate system[18]
$$
ds ^2 = e ^{2( \gamma - \psi )}(dt ^2 - dr_{_c} ^2 ) -
e ^{2 \psi }dz ^2 -  {\tilde \alpha} ^2 e ^{-2 \psi } d \phi ^2 
\eqno (2.8)
$$
(where $\gamma, \ \psi, {\tilde \alpha}$ are independent of $z, \phi$) with the
string energy-momentum tensor as the source: $$
T_{\mu\nu} = 2\eta^2 \nabla_\mu X\nabla_\nu X + 2\eta^2X^2P_\mu P_\nu
 -{2\beta \over \lambda } F_{\mu\sigma}F_\nu^{\ \sigma} - {\cal L}g_{\mu\nu}
\eqno (2.9) 
$$
in unrescaled coordinates. To rescale the coordinates, we set
$R = \sqrt{\lambda}\eta r_{_c}$, $\alpha = \sqrt{\lambda}\eta {\tilde
\alpha}$, and for future comparison, we write the rescaled version of the
energy and stresses ${\hat T}_{ab} = T_{ab}/(\lambda\eta^4)$ 
$$
\eqalignno{
{\hat T}^0_0 &= {\cal E} = e ^{-2(\gamma-\psi)} X ^{\prime 2} + {e^{2\psi} X^2
P^2 \over {\beta^{-1}}{\alpha} ^2} + e^{2(\gamma-2\psi)} P ^{\prime 2}  / {\alpha} ^2 +
(X^2 -1) ^2 / 4  & (2.10a) \cr 
{\hat T}^R_R &= - {\cal P}_R = - e ^{-2(\gamma-\psi) } X ^{\prime 2} +
{e^{2\psi} X^2 P^2 \over {\beta^{-1}}{\alpha} ^2} - e^{2(\gamma-2\psi)}P ^{\prime 2}  /
{\alpha} ^2 + (X^2 -1) ^2 / 4 & (2.10b) \cr 
{\hat T}^\phi_\phi&=  -{\cal P}_\phi = e ^{-2(\gamma-\psi) } X ^{\prime 2} -
{e^{2\psi} X^2 P^2 \over {\beta^{-1}} {\alpha} ^2} -e^{2(\gamma-2\psi)} P ^{\prime 2}  /
{\alpha} ^2 + (X^2 -1) ^2 / 4 \;\;\; & (2.10c) \cr 
{\hat T}^z_z &= - {\cal P}_z = {\hat T}^0_0 .& (2.10d)\cr}
$$
Also for future reference, the Bianchi identity gives
$$
{\cal P}' _{R} + ({\cal P} _{R} -  {\cal P}  _{\phi }) ({\alpha '\over \alpha}
- \psi') + \gamma '{\cal P} _R + \gamma '{\cal E}= 0. \eqno (2.11)
$$

To zeroth order (flat space)
$$
\alpha = R \;\; , \;\;\; \psi=\gamma=0 \;\; , \;\;\; X=X_0\;\;,\;\;\; P=P_0,
\eqno (2.12)
$$
and (2.11) gives
$$
(R{\cal P}_{_0R})' = {\cal P}_{_0\phi}
\eqno (2.13)
$$

To first order in $\epsilon=8\pi G\eta^2$ the string metric is given by[19]
$$
{\alpha} = \left [1- \epsilon \int _0 ^R
R ({\cal E}_{_0} - {\cal P}_{_0R})d R \right ] R +
\epsilon \int _0 ^R  R ^2 ({\cal E}_{_0} - {\cal P}_{_0 R})dR,
\eqno (2.14a)
$$
$$
\gamma = 2\psi= \epsilon \int _0 ^R R {\cal P}_{_0 R} dR .
\eqno (2.14b)
$$
Where the subscript zero indicates evaluation in the flat space limit.
Note that when the radial stresses
do not vanish, there is a scaling between the
time, $z$ and radial coordinates for an observer at infinity and those for an
observer sitting at the core of the string. The only case in which these
stresses do vanish is when ${\beta}=1$. In this case
the field equations reduce to 
$$
\eqalign{
X' &= XP/\alpha \cr
P' &= {\half} \alpha (X^2 -1) \cr
\alpha ' &= 1 - \epsilon [ (X^2-1)P +1] \cr
\gamma&=\psi=0 \cr
}
\eqno(2.15)
$$
a first order set of coupled differential equations as one might expect from
the fact that the solution is supersymmetrizable.

We conclude this section by demonstrating the asymptotically conical nature of
the corrected metric[19,20] (see also [21-24] for discussions involving model
cores rather than vortices). Note that since the string functions $X$ and $P$
rapidly fall off to their vacuum values outside the core, the integrals 
in (2.14)
rapidly converge to their asymptotic, constant, values. Let
$$
\epsilon \int _0 ^R
R ({\cal E}_{_0} - {\cal P}_{_0R})d R = A, \;\;\;\;
\epsilon \int _0 ^R R^2 ({\cal E}_{_0} - {\cal P}_{_0R})d R = B
\;\;\;\;\;{\rm and} \;\;\;\;\;
\epsilon \int _0 ^R R {\cal P}_{_0 R} = C
\eqno (2.16)
$$
then the asymptotic form of the metric is
$$
\eqalign{
ds^2 &= e^C[dt^2-dr_c^2-dz^2] - r_c^2 (1-A+B/r_c)^2 e^{-C} d\phi^2\cr
&= d{\hat t}^2 - d{\hat r}_c^2 - d{\hat z}^2 - {\hat r}_c^2 (1-A)^2 e^{-2C}
d\phi^2 \cr}
\eqno (2.17)
$$
where ${\hat t} = e^{C/2}t,\ {\hat z} = e^{C/2}z$,  and ${\hat r}_c =
e^{C/2}(r_c + B/(1-A))$.  This is seen to be conical with a deficit
angle 
$$
\Delta = 2\pi (A+C) = 2\pi \epsilon\int R {\cal E}_{_0} dR
= 16 \pi^2 G \int r_c T^0_0 dr_c = 8\pi G\mu
\eqno (2.18)
$$
where $\mu$ is the energy per unit length of the string.
Notice that the deficit angle is independent of the radial stresses,
but that there is a red/blue-shift of time between infinity and the
core of the string if they do not vanish.

\vskip 2mm

\noindent{\bf 3. String in background Schwarzschild metric.}

In solving on a Schwarzschild background there are two coordinate systems
 we could consider.

i) Spherical (Schwarzschild) coordinates
$$
ds^2 = \left ( 1-{2GM\over r_s} \right ) dt^2 - \left ( 1-{2GM\over r_s}
\right )^{-1} dr_s^2 - r_s^2 (d\theta^2+\sin^2\theta d\phi^2)
\eqno (3.1)
$$
This is the usual Schwarzschild metric. This coordinate system is good for
analyzing the existence of a background solution, but, not being tailored to
the symmetries of the full problem, it does not deal with gravitational
back-reaction well.

ii) Axisymmetric (Weyl) coordinates[25]

\noindent Here
$$
ds ^2 ={\textstyle{R_1+R_2-2GM\over R_1+R_2+2GM}} dt ^2 -
{\textstyle{(R_1+R_2+2GM)^2\over 4R_1R_2}} (dz ^2 +dr_c ^2 ) -
r_c^2 {\textstyle{(R_1+R_2+2GM)\over (R_1+R_2-2GM )}} d \phi^2 
\eqno (3.2)
$$
where
$$
R_1^2 = (z-GM)^2 + r_c^2 \;\;\; , \;\;\;\; 
R_2^2 = (z+GM)^2 + r_c^2 
\eqno (3.3)
$$
The transformation between the two systems is given by
$$
z=(r_s-GM)\cos\theta \;\; , \;\;\; r_c^2 = r_s(r_s-2GM) \sin^2\theta
\eqno (3.4)
$$
Weyl coordinates are appropriate for analyzing the gravitational back
reaction, however are rather cumbersome for the background solution problem.

The rest of this section is devoted to arguing existence of a vortex solution 
in a black hole background. Formally, this means taking the somewhat 
artificial limit $G\eta^2 \to 0$ keeping $GM$ fixed. It is straightforward
to show that it is consistent to take
$$
P_\mu = P \nabla_\mu \phi
\eqno(3.5)
$$
and letting
$$
\eqalign{
r=\sqrt{\lambda}\eta r_{_s} \cr
E = \sqrt{\lambda}\eta GM \cr
}
\eqno (3.6)
$$
in the Schwarzschild metric, the equations of motion for $X$ and $P$ are
$$
\eqalignno{
- {1\over r^2} \partial_r [ r(r-2E)\partial_r X] - {1\over r^2\sin\theta}
\partial_\theta [ \sin\theta\partial_\theta X ] + \half X(X^2-1) + 
{XP^2 \over r^2\sin^2\theta} &=0 & (3.7a) \cr
\partial_r [ (1-2E/r)\partial_r P ]+ {\sin\theta\over r^2}\partial_\theta
[ \csc\theta\partial_\theta P]  - {\beta^{-1}} X^2 P &=0 & (3.7b)  \cr
}
$$
We proceed at first by taking a ``thin string limit", in other words we
assume $2E \gg 1$. This is equivalent to requiring $M\gg 1000$ Kg for the
parameters of the GUT string. We will then consider thicker and higher winding
number strings.

First of all, let us try
$$
X = X(r\sin\theta), P=P(r\sin\theta)
\eqno (3.8)
$$
substituting these forms into (3.7), and writing (suggestively)
$r\sin\theta=R$
we get
$$
\eqalignno{
- X'' - X'/R + \half X(X^2-1) + XP^2/R^2 + 2E\sin^2\theta/r [ X''+X'/R] &=0
& (3.9a) \cr
P''-P'/R - {\beta^{-1}} X^2P + 2E\sin^2\theta/r[-P''+P'/R] &=0 & (3.9b) \cr
}
$$
where prime denotes derivative with respect to $R$. 
If we were to input $X,P = X_0, P_0$ then these equations would be satisfied
up to errors of the form
$$
{2E\sin^2\theta\over r}\times[{\hbox {\rm other terms in equation}}]
\eqno (3.10)
$$
However, since $r\sin\theta=R\sim O(1)$ in the core of the string, 
$\sin\theta = O(1/r) $, and the errors are $O(E/r^3) < O(1/E^2) \ll 1$.
Thus the Nielsen-Olesen solution is in fact a good solution throughout
and beyond the core of the string, whether near the event horizon or not.
By the time the premultiplying term in the errors is significant, we are well
into the exponential fall-off of the vortex, and essentially in vacuum.
However, since we are interested in showing that the event horizon can support
the vortex, for completeness we include the solution to order $1/E$
on the horizon: 
$$ 
1-X \sim e^{-2E\theta} + e^{-2E(\pi-\theta)}, \;\;\;
P \sim  e^{-2E\sqrt{{\beta^{-1}}}\theta} + e^{-2E\sqrt{{\beta^{-1}}}(\pi-\theta)}
\eqno (3.11)
$$
Of course, these are only approximate forms, and do not 
prove existence of a solution to (3.7). However, they will provide a good
approximation to the true solution, if such can be shown to exist. We will
provide such evidence in the form of numerical solutions later in this
section. For the moment, we conclude this description of the thin string limit
by examining the equations of motion in the extended Schwarzschild spacetime
in terms of Kruskal coordinates.

One problem with using Schwarzschild coordinates for our analysis is that they
are singular on the event horizon. This is, of course, purely a coordinate
singularity, but since demonstrating a dressing of the event horizon is central
to this paper, we will examine the thin vortex solution in coordinates that are
not singular at the event horizon, namely Kruskal coordinates, to convince the
reader that the analytic approximation really does hold true at  the event
horizon. Kruskal coordinates are based on the incoming and outgoing radial null
congruences of the Schwarzschild spacetime, but we shall instead use a Kruskal
`time' and `space' coordinate defined as follows.
$$
\eqalign{
T &= (r_s-2GM)^{1/2} e^{r_s/4GM} \sinh \left ( {t\over 4GM} \right ) \cr
Y &= (r_s-2GM)^{1/2} e^{r_s/4GM} \cosh \left ( {t\over 4GM} \right ) \cr
}
\eqno (3.12)
$$
In terms of which the metric is
$$
ds^2 = {16 G^2M^2 \over r_s} e^{-r_s/2GM} (dT^2 - dY^2) - r_s^2 (d\theta^2 +
\sin^2\theta d\phi^2)
\eqno (3.13)
$$
This metric is clearly regular away from $r_s=0$, and the future and past event
horizons are represented by $T=Y>0$ and $-T=Y>0$ respectively. What we would like
to show is that $X=X_0(r\sin\theta)$ and $P=P_0(r\sin\theta)$ are good
solutions (expressed in Kruskal coordinates) in the vicinity of the horizon,
$|T|=|Y|$.

First note that near the horizon
$$
r = \sqrt{\lambda}\eta r_s = 2E - e^{-1}\sqrt{\lambda}\eta (T^2-Y^2) +
O(T^2-Y^2)^2
\eqno (3.14)
$$
where here $e=2.718...$ is the natural number. This implies that 
$$
\partial_T X = {-2T\sqrt{\lambda}\eta \over e}\sin\theta X'(R) \;\;\; ; \;\;\;\;
\partial_Y X = {2Y\sqrt{\lambda}\eta \over e}\sin\theta X'(R).
\eqno (3.15)
$$
and hence
$$
\left[{1\over \sqrt{g}}\partial_T g^{TT} \sqrt{g} \partial_T +
{1\over \sqrt{g}}\partial_R g^{RR} \sqrt{g} \partial_R \right ] X
= \lambda \eta^2 \left [ {(2E-r)\over 2E} \sin^2\theta X''(R) + {(E-r)\over
2E^2} X'(R)\sin\theta \right ] 
$$
$$
\left [ {1\over \sqrt{g}}\partial_T {g^{TT}\sqrt{g}\over
r_s^2\sin^2\theta}\partial_T + {1\over \sqrt{g}}\partial_R { g^{RR} \sqrt{g}
\over r_s^2\sin^2\theta}  \partial_R \right ] P = \lambda\eta^2
\left [{(2E-r) \over 2E^3} P'' - {P'\over E^2 R}\right]
\eqno (3.16)
$$
In other words, as in Schwarzschild coordinates, the Nielsen-Olesen solution
solves the equations of motion to O($E^{-2}$) near, on, and even beyond the
event horizon. Indeed, replacing 
$$
r = \sqrt{\lambda}\eta {r^*}^{-1} \{ 2GM \log(Y^2-T^2)\},
\eqno (3.17)
$$
where $r^*(r_s)$ is the tortoise coordinate, indicates that the approximation
holds true well within the event horizon for larger black holes.

Having established that it is possible for the horizon
to support $N=1$ vortices, we now turn to the large $N$
case, where analytic approximations are available
[26]. First of all, note that  a string with $N\gg1$
 has a core radius of order $\sqrt N$. To see this, consider the rather
special case of $\beta=1$ (for general $\beta$, see [26]). Setting
$X=\xi^N$, we have from (2.15)
$$
NP' = {\half} R (\xi^{2N}-1) \hskip 2cm 
{\xi'\over\xi} = {P\over R}
\eqno (3.18)
$$ 
in the absence of gravity. These are in fact the same as the large-N
equations for general $\beta$, it is in the sub-leading terms that the
two cases differ. These are solved in the core by
$$
P = 1 - {R^2 \over 4N} \;\;\; ; \;\;\; \xi \propto Re^{-R^2/2N}
\eqno (3.19)
$$
using $0<\xi<1$. 
The transition to vacuum, and hence a different
approximate solution to the above can be seen to occur quite abruptly
at $R=O(\sqrt{N})$.
Thus the
condition that the string be thin compared with the black
hole is now  $2E\gg\sqrt N$, and in that case the previous
arguments still apply, since 
$$
{2E \sin^2 \theta \over r}
 = {2ER^2 \over r^3} < { 2EN\over r^3} \leq {N \over
(2E)^2} \ll 1 \ \ ,
$$
i.e., the Nielsen Olesen solution for a large-N string is
good also on the event horizon.

Now consider the opposite limit, where the string is much
bigger than the black hole, $\sqrt N \gg 2E \ $. The black hole sits well
inside the core of the string, in a region where $X \approx 0, \ P
\approx 1 - pR^2$.
In the absence of a black hole the
P-equation simply states that the magnetic field
is constant throughout the core of the string. Notice
that since we are ignoring $X^2$ in the large-N expansion, (3.9b) implies
that the presence of the  black hole  does not affect  the
large-$N$ $P$-equation, so we still find
$ P (R) \sim 1 - pR^2 $.
The magnetic field will still be
constant (and equal to $-2p$) in the string core.
However, its value may change due to the black hole.
Notice that we may expect a slight ``squeezing'' of the
string core  due to the black hole. To
see this, consider rewriting the $P$-equation (3.9b) in the form
$$
P'' - ({P'/ R}) - {\beta^{-1}} (r, \theta) P X^2 = 0. 
\eqno (3.20)
$$
On the equatorial plane of the black hole, $\sin\theta = 1$, 
and ${\beta^{-1}}(R) = {\beta^{-1}}
/ (1 - {2E/ R})$. For $R\gg 2E$ we have the vacuum
solution $X=1$, $P=0$. As we come in towards the horizon,
$P$ has to leave its vacuum value - however, the effective
value of ${\beta^{-1}}$ (which measures an ``effective mass'' for
$P$) is increasing. Compared with the situation where
there is no black hole, $P$ should be  more reluctant to
leave its vacuum value. The magnetic field will remain
zero for as long as possible and, as a result, the string
core is somewhat smaller around the black hole. Note that
this argument does not  apply for global strings,  where
 numerical simulations indeed show that it is a much smaller
effect.

Now consider the $X$ equation, to leading
order in $N$ \quad $( r > 2E)$:
$$ 
\biggl( 1 - {2E \sin^2 \theta \over
r}\biggr) \biggl({\xi'\over\xi}\biggr)^2 = 
\biggl( 1 - {2E R^2 \over
( R^2 + z^2)^{3/2}}\biggr) \biggl({\xi'\over\xi}\biggr)^2 =
{P^2\over R^2} +
O\left({1 \over N^2}\right)  
\eqno (3.21)
$$ 
We want to show that the solutions to
this equation are regular at the horizon. The equation
becomes singular at $\sin\theta = 1,\  r = 2E$, (or
$z=0, \ R = 2E$),  the equatorial
plane of the horizon, so let us  integrate  the equations 
to leading order in $1/N$ and $z/R$:
$$
{\xi'\over\xi}\sqrt {\biggl( 1 - {2E  \over R}\biggr)}  = {P\over R} \ \ .
\eqno (3.22)
$$
We find
$$
\xi = K \biggl[ R-E + \sqrt {R(R-2E)} \biggr] ^{(1 - {3\over 2}
p E^2)}  e^{ -p \bigl[ {3E + R \over 2} \bigr] \sqrt{
R(R-2E)}} \ \ ,
\eqno (3.23)
$$ 
which is finite at R=2E. 
The constant $K$ can be fixed by the requirement
that $\xi \sim 1 $ when $P \sim 0$, i.e., at $R=1/\sqrt
p$.

Finally, note that the horizon seems to be capable of
supporting global strings as well, in spite of the fact
that we do not expect to be able to account for gravitational back
reaction consistently, since the energy per unit length of
a global string diverges and the self-gravitating global string
spacetime is singular[27,28]. To find the global string
solution in the black hole background we simply set $P=1$
everywhere, to find
$$
- X'' - X'/R + \half X(X^2-1) + N X/R^2 +
2E\sin^2\theta/r [ X''+X'/R] =0 
\eqno (3.24)
$$
The case where the global string is thin compared to the
black hole works as before - the vortex is essentially
undisturbed. In the case where the string is bigger that
the black hole we can take $p \to 0$ in (3.23), so the
solution is again  regular at the horizon. We
conclude that the presence  of the black
hole is, if anything, less noticeable than in  the local
string case, as can be seen in the numerical simulations
described next.

 \bigskip\bigskip\noindent {\it
Numerical results}

We will now provide confirmation of the previous analytic
arguments  by means of a numerical solution
of the equations of motion outside the event horizon. To
this end, we note that the equations are elliptic outside
the event horizon, parabolic on it and hyperbolic inside
it. Some care is therefore required with specification of
the boundary conditions.

At large radii we want to recover the NO solutions, while the symmetry axes
outside the horizon must form the core of the string:
$$
(X,P)=\cases{(1,0),&$r\to\infty$;\cr (0,1),&$r\ge2E$, $\theta=0$,
$\pi$.}
\eqno(3.25)
$$ 
On the horizon the equation turns parabolic, taking the form
$$
\eqalignno{
\left.{1\over 2E}\partial_r X\right|_{r=2E}
&=-{1\over4E^2\sin\theta}\partial_\theta[ \sin\theta\partial_\theta X]+\half
X(X^2-1)+{N^2XP^2\over4E^2\sin^2\theta},&(3.26a)\cr
-\left.{1\over2E}\partial_r P\right|_{r=2E}
&={\sin\theta\over4E^2}\partial_\theta [ \csc\theta\partial_\theta P] 
-{\beta^{-1}}
X^2P.&(3.26b)
} 
$$
A given choice of $X(2E,\theta)$ and $P(2E,\theta)$ on the horizon therefore
directly implies $\partial_r X(2E,\theta)$ and $\partial_r P(2E,\theta)$.
Generally boundary values and normal derivatives 
cannot be specified simultaneously when solving an elliptic equation:
either suffices to determine the solution completely. 
A particular choice of $X$ and $P$ on the horizon and the
corresponding normal derivatives will not normally generate the same solution
outside the horizon. The boundary conditions on the event horizon are
thus determined by the requirement that both solutions coincide.

A practical algorithm for solving the equations of motion in a 
Schwarzschild background numerically then is 
as follows. We employ a uniformly spaced polar grid $\{(r_i,\theta_j)\}$,
with boundaries at $r=2E$ and a large radius $r_L\gg2E$, and $\theta$ 
ranging from 0 to $\pi$. Then we approximate the derivatives with
finite-difference expressions on the grid. Writing $F_{00}$ for the value of
the field $F$ at the grid point $(r_i,\theta_j)$, and similarly 
$F_{\pm0}$ for $F(r_{i\pm1},\theta_j)$ and $F_{0\pm}$ for
$F(r_i,\theta_{j\pm1})$, we obtain the finite difference equations 
$$
X_{00}={
{2\over r}\left(1-{E\over r}\right){X_{+0}-X_{-0}\over2\Delta r} 
+{\cot\theta\over r^2}{X_{0+}-X_{0-}\over2\Delta\theta}
+\left(1-{2E\over r}\right){X_{+0}+X_{-0}\over \Delta
r^2} +{X_{0+}+X_{0-}\over r^2\Delta\theta^2}
\over \left(1-{2E\over r}\right){2\over\Delta r^2}+
{2\over r^2\Delta\theta^2}  +\half(X_{00}^2-1)+
\left(NP_{00}\over r\sin\theta\right)^2},
\eqno(3.27a)
$$
$$
P_{00}={
 \left(2E\over r^2\right){P_{+0}-P_{-0}\over2\Delta r}
-\cot\theta{P_{0+}-P_{0-}\over2\Delta\theta r^2} 
+\left(1-{2E\over r}\right){P_{+0}+P_{-0}\over \Delta
r^2}+{P_{0+}+P_{0-}\over r^2\Delta\theta^2} \over
\left(1-{2E\over r}\right){2\over\Delta r^2}+
{2\over r^2\Delta\theta^2}+{\beta^{-1}} X_{00}^2}
\eqno(3.27b) 
$$ 
inside the grid, and
$$
\eqalignno{
X_{00}=&
{{EX_{+0}\over\Delta r}
+{X_{0+}+X_{0-}\over2\Delta\theta^2}
+\cot\theta{X_{0+}-X_{0-}\over4\Delta\theta}\over
{E\over\Delta r}+{1\over\Delta\theta^2}+E^2(X_{00}^2-1)+
   {1\over2}\left(NP_{00}\over\sin\theta\right)^2}, & (3.27c)
\cr
P_{00}=&
{{EP_{+0}\over\Delta r}
+{P_{0+}+P_{0-}\over2\Delta\theta^2}
-\cot\theta{P_{0+}-P_{0-}\over4\Delta\theta}\over
{E\over\Delta r}+{1\over\Delta\theta^2}
+2E^2{\beta^{-1}} X_{00}^2} & (3.27d)\cr 
}
$$
on the horizon.

Initial values for $X$ and $P$ are assigned on the boundaries
according to (3.25); on the horizon we initially set $X=0$, $P=1$. $X$
and $P$ are then iteratively adjusted on the interior grid points
according to (3.27a,b), analogous to the Gauss-Seidel scheme for
linear elliptic equations [29]. After each pass
through the interior grid points, the $r$-gradients of $X$ and $P$
just outside the horizon are calculated, and equations (3.27c,d)
iterated to derive new values for $X$ and $P$ on the horizon (for given
$r$-gradients, the equations on the horizon are one-dimensional
elliptic equations). The
whole process is then iterated to convergence. In order to speed up
convergence, the grid is over-relaxed: instead of replacing $X$ and $P$ by
the right-hand sides of eqs.\ (3.27), $wX_{new}+(1-w)X$ with $1<w<2$
is used. The optimal value for the over-relaxation parameter $w$ is
found by trial and error, and depends on the number of grid points and
on the differential equation.

Sample results are presented in Figures 1--7, and confirm the analytic
arguments above. Figures 1--4 show a sequence of solutions
with increasing winding number (and therefore string thickness)
threading an E=10 black hole. Qualitatively, the string simply
continues regardless of the black hole, though some mild pinching of
the magnetic flux does take place.  Figures 5 and 6 compare a local and
global string with the same winding number and `width'; the global string
is apparently fatter due to the power law, as opposed to exponential, 
fall-off in the fields. Figure 7 shows a comparison between the 
numerically obtained solutions and the Nielsen-Olesen analytic
approximation that will be used in the next section. As can be seen, 
even for a string of non-negligible thickness, this is still an excellent
approximation.

\vskip 2mm

\noindent{\bf 4. Gravitating strings}

In order to get the gravitational effect of the string superimposed on the
black hole, we need to consider a general static axially symmetric metric
$$
ds ^2 = e ^{2 \psi }dt ^2 
-e ^{2( \gamma - \psi )}(dz ^2 + dr_c ^2 ) -
\tilde{\alpha}^2 e ^{-2 \psi } d \phi ^2
\eqno (4.1)
$$
where $\psi, \gamma,\tilde{\alpha}$ are independent of $t, \phi$. Notice that
this is related to (2.8) through $z\to it, \ t\to iz$. We then apply an
iterative procedure to solving the field equations, starting with the
background solutions (3.2) and the \N-\O~forms of $X$ and $P$, and  expanding
the equations of motion in terms of $\epsilon=8\pi G\eta^2$ which is assumed
small. ($\epsilon \leq 10^{-6}$ for GUT strings.) We first rescale coordinates
to bring them into line with the rescaled  Schwarzschild coordinates used in
the previous section $$ \eqalign{
\rho &= \sqrt{\lambda}\eta r_c \cr
\zeta &= \sqrt{\lambda}\eta z \cr
\alpha &= \sqrt{\lambda}\eta\tilde{\alpha} \cr
}
\eqno (4.2)
$$
and rewrite
$$
\eqalign{
{\hat R}_1^2 &= (\zeta-E)^2 + \rho^2 \cr
{\hat R}^2_2 &= (\zeta+E)^2 + \rho^2 \cr
}
\eqno (4.3)
$$
In terms of the rescaled coordinates and energy-momentum tensor, the
Einstein equations become
$$
\eqalignno{
\alpha_{,\zeta\zeta} + \alpha_{,\rho\rho} &= -\epsilon \sqrt{-g}
({\hat T} ^\zeta_\zeta + {\hat T}^\rho_\rho) & (4.4a) \cr
(\alpha\psi_{,\zeta})_{,\zeta} + (\alpha \psi_{,\rho})_{,\rho} &=
{\half}\epsilon \sqrt{-g}({\hat T} ^0_0 - {\hat T} ^\zeta_\zeta
-{\hat T}^\rho_\rho - {\hat T}^\phi_\phi) & (4.4b) \cr
(\alpha_{,\zeta}^2 + \alpha_{,\rho}^2) \gamma_{,\rho} &=
\epsilon \sqrt{-g} ( \alpha_{,\rho} {\hat T}^\zeta_\zeta - \alpha_{,\zeta}
{\hat T}^\rho_\zeta) + \alpha\alpha_{,\rho}(\psi_{,\rho}^2 - \psi_{,\zeta}^2)
\cr & + 2\alpha\alpha_{,\zeta} \psi_{,\zeta} \psi_{,\rho} + \alpha_{,\rho}
\alpha_{,\rho\rho} + \alpha_{,\zeta} \alpha_{,\zeta\rho} & (4.4c) \cr
(\alpha_{,\zeta}^2 + \alpha_{,\rho}^2) \gamma_{,\zeta} &= -
\epsilon \sqrt{-g} ( \alpha_{,\zeta} {\hat T}^\zeta_\zeta - \alpha_{,\rho}
{\hat T}^\rho_\zeta) - \alpha\alpha_{,\zeta}(\psi_{,\rho}^2 - \psi_{,\zeta}^2)
\cr & + 2\alpha\alpha_{,\rho} \psi_{,\zeta} \psi_{,\rho} - \alpha_{,\zeta}
\alpha_{,\rho\rho} + \alpha_{,\rho} \alpha_{,\zeta\rho} & (4.4d) \cr
\gamma_{,\rho\rho} + \gamma_{,\zeta\zeta} &= - \psi_{,\rho}^2
- \psi_{,\zeta}^2 - \epsilon e^{2(\gamma-\psi)} {\hat T}^\phi_\phi
& (4.4e) \cr
}
$$
where the energy momentum tensor is given by
$$
\eqalign{
{\hat T} ^0_0 &= V(X) + {X^2P^2\over \alpha^2e^{-2\psi}} 
+ \left [ {(P_{,\rho}^2 + P_{,\zeta}^2) \over {\beta^{-1}} \alpha^2 e^{-2\psi}} 
+ (X_{,\rho}^2 + X_{,\zeta}^2) \right ]  e^{-2(\gamma-\psi)} \cr
{\hat T}^\phi_\phi &= V(X) - {X^2P^2\over \alpha^2e^{-2\psi}}
+ \left [ - {(P_{,\rho}^2 + P_{,\zeta}^2) \over {\beta^{-1}} \alpha^2 e^{-2\psi}} 
+ (X_{,\rho}^2 + X_{,\zeta}^2) \right ]  e^{-2(\gamma-\psi)} \cr
{\hat T}^\rho_\rho &=  V(X) + {X^2P^2\over \alpha^2e^{-2\psi}} 
+ \left [ {(P_{,\zeta}^2 - P_{,\rho}^2) \over {\beta^{-1}} \alpha^2 e^{-2\psi}} 
+ (X_{,\zeta}^2 - X_{,\rho}^2) \right ]  e^{-2(\gamma-\psi)} \cr
{\hat T}^\zeta_\zeta &=  V(X) + {X^2P^2\over \alpha^2e^{-2\psi}} 
- \left [ {(P_{,\zeta}^2 - P_{,\rho}^2) \over {\beta^{-1}} \alpha^2 e^{-2\psi}} 
+ (X_{,\zeta}^2 - X_{,\rho}^2) \right ]  e^{-2(\gamma-\psi)} \cr
{\hat T}^\rho_\zeta &= - 2 e^{-2(\gamma-\psi)} \left [ X_{,\zeta} X_{,\rho}
+ {P_{,\zeta} P_{,\rho} \over {\beta^{-1}} \alpha^2 e^{-2\psi}} \right ] \cr
}
\eqno (4.5)
$$

We now write $\alpha = \alpha_0+\epsilon\alpha_1$ etc., and solve the Einstein
equations (4.4), and the string equations (2.4), iteratively. To zeroth order,
we have the background solutions
$$
\eqalignno{
&\alpha_0 = \rho \;\;\; , \;\;\; 
\psi_0 = {\half}\log{\textstyle{R_1+R_2-2GM\over
R_1+R_2+2GM}} \;\;\;,\;\;\; \gamma_0 =
{\half}\log{\textstyle{(R_1+R_2-2GM)(R_1+R_2+2GM)\over
4R_1R_2}}  & (4.6a) \cr &X=X_0(R)
\;\;\;\;\;\;\;,\;\;\;\;\;\;\;\;\;P=P_0(R) \ \ , & (4.6b)
\cr} 
$$ 
where $R_1$ and $R_2$ were defined in (3.3). In these
coordinates $R=r\sin\theta=\rho e^{-\psi_0}$, so (4.6b)
indicates that many of the terms in ${\hat T}^a_b$ are simply
functions of $R$. 

Before proceeding to calculate the back reaction
however, it is prudent to check that the energy-momentum tensor (4.5)
will admit a geodesic shear free event horizon. Recall that we require
$T^0_0 - T^r_r=0$ on the horizon in Schwarzschild coordinates. This is
clearly satisfied at $\theta=0$, where the energy and tension balance,
but what about $\theta\neq0$? In Weyl coordinates, this corresponds
to ${\hat T}^0_0 - {\hat T}^\rho_\rho=0$ for $\rho\to 0$, $\zeta\neq\pm E$.
From (4.5) we see that this is given by
$$
\eqalign{
{\hat T}^0_0 - {\hat T}^\rho_\rho&= 2 e^{2(\gamma-\psi)} \left [
{\beta P_{,\rho}^2 \over \alpha e^{-2\psi} } + X_{,\rho}^2 \right ] \cr
&= {8 {\hat R}_1 {\hat R}_2 \over ({\hat R}_1+{\hat R}_2 +2E)^2}
\left ( {dR\over d\rho} \right )^2 \left [ {\beta P'(R)^2\over R^2}
+ X'(R)^2 \right ] \cr
}
\eqno (4.7)
$$
All terms in this expression remain finite and non-zero as $\rho\to0$ ($R\to0$) 
except for ${dR\over d\rho}$. Using the transformation (3.4)
between the Schwarzschild and Weyl coordinates, we have 
$$
\eqalign{ {\partial R\over \partial \rho} &= {\rho (r
-E\sin^2\theta) \over {\hat R}_1{\hat R}_2 \sin\theta} \cr
{\partial R\over \partial \zeta} &= 
-{Er \sin\theta\cos\theta \over {\hat R}_1{\hat R}_2}
\cr}\eqno (4.8)
$$
hence ${dR\over d\rho}\to0$ as $\rho\to 0$ and 
${\hat T}^0_0 - {\hat T}^\rho_\rho$ is indeed zero on 
the horizon. Thus there is no
gravitational obstruction, at least in this linearized method, of painting
the vortex onto the horizon.

Proceeding with the analysis of the energy momentum tensor,
we use ${\hat R}_2^2 - {\hat R}_1^2 = 4E\zeta$, which gives
${\hat R}_1{\hat R}_2 = E^2\sin^2\theta + r(r-2E)$,
thus near the core of the string, where $\sin\theta = O(E^{-1})$
$$
R^2_{,\zeta} + R^2_{,\rho} =  {r^2\over {\hat R}_1{\hat R}_2} 
\left ( 1 - {2E\over r}\sin^2 \theta \right ) 
\sim {r^2\over {\hat R}_1{\hat R}_2} = e^{2(\gamma_0-\psi_0)}.
\eqno (4.9)
$$

Therefore, in and near the core of the string the zeroth order rescaled
energy-momentum tensor now reads 
$$
\eqalign{
{\hat T} ^0_{o0} &= V(X_0) + {X_0^2P_0^2\over R^2} 
+ {P_0^{\prime 2} \over {\beta^{-1}} R^2 }
+ X_0^{\prime 2} + O(E^{-2}) \cr
{\hat T}^\phi_{o\phi} &= V(X_0) - {X_0^2P_0^2\over R^2}
- {P_0^{\prime 2} \over {\beta^{-1}} R^2 }
+ X_0^{\prime 2} + O(E^{-2}) \cr
{\hat T}^\rho_{o\rho} &=  V(X_0) + {X_0^2P_0^2\over  R^2}
+ \left [ {P_0^{\prime 2} \over {\beta^{-1}} R^2 } + X_0^{\prime 2} 
\right ] ( R_{,\zeta}^2 - R_{,\rho}^2) e^{-2(\gamma_{_o}-\psi_{_o})} \cr
{\hat T}^\zeta_{o\zeta} &=  V(X_0) + {X_0^2P_0^2\over  R^2}
-  \left [ {P_0^{\prime 2} \over {\beta^{-1}} R^2 } + X_0^{\prime 2} 
\right ] ( R_{,\zeta}^2 - R_{,\rho}^2) e^{-2(\gamma_{_o}-\psi_{_o})} \cr
{\hat T}^\rho_{o\zeta} &= - 2 e^{-2(\gamma_{_o}-\psi_{_o})} 
R_{,\zeta}R_{,\rho} \left [ {P_0^{\prime 2} \over {\beta^{-1}} R^2 } 
+ X_0^{\prime 2} \right ] + O(E^{-2})\cr
}
\eqno (4.10)
$$
and the combinations used in equations (4.4a,b,e) are all purely
functions of $R$. This strongly suggests looking for metric perturbations
as functions of $R$. However, we must check that the left hand sides of
these equations can be written as appropriate functions of $R$.

Consider $R=\rho e^{-\psi_{_o}}$, 
$$
R_{,\zeta} = -R\psi_{_o,\zeta} \hskip 3mm  \Rightarrow \hskip 3mm
R_{,\zeta\zeta} =  -R \psi_{_o,\zeta\zeta} + { R_{,\zeta}^2\over R }
\eqno (4.11) 
$$
and
$$
R_{,\rho} = ({1\over \rho} - \psi_{_o,\rho} ) R \hskip 3mm \Rightarrow \hskip
3mm  R_{,\rho\rho} = - (\rho\psi_{_o,\rho})_{,\rho} {R\over\rho} - R_{,\rho}
( {1\over\rho} - {R_{,\rho }\over R})
\eqno (4.12)
$$
imply 
$$
R_{,\zeta\zeta} + R_{,\rho\rho} = {R_{,\zeta}^2+ R_{,\rho }^2\over R}
-  {R_{,\rho }\over \rho} - [(\rho\psi_{_o,\rho})_{,\rho} + 
\rho \psi_{_o,\zeta\zeta} ]{R\over \rho} 
= -{E\sin\theta \over r{\hat R}_1{\hat R}_2} =
O(E^{-3}) e^{2(\gamma_{_o} - \psi_{_o})} 
\eqno (4.13)
$$
where we have used the zeroth order equation of motion for $\psi_0$.
Therefore 
$$
\partial_\zeta^2 + \partial_\rho^2 = (R^2_{,\zeta}+R^2_{,\rho})
{d^2\over dR^2} + (R_{,\zeta\zeta} + R_{,\rho\rho} ) {d\over dR}
= e^{2(\gamma_{_o} - \psi_{_o})} \left [ {d^2\over dR^2} + O(E^{-3}) \right ]
\eqno (4.14)
$$
in the core of the string. Exterior to the core, the vacuum equations 
will apply.
We now solve (4.4a,b,e) to first order in $\epsilon$, namely
$$
\eqalignno{
\alpha_{1,\zeta\zeta} + \alpha_{1,\rho\rho} &= -2\rho e^{2(\gamma_0-\psi_0)} 
(V(X_0)+{X_0^2P_0^2\over R^2}) \cr
& = -\rho e^{2(\gamma_0-\psi_0)}({\cal E}_0 -
{\cal P}_{_{0R}}) & (4.15a) \cr
\alpha_{1,\zeta} \psi_{0,\zeta} + \alpha_{1,\rho} \psi_{0,\rho} +
\rho \psi_{1,\zeta\zeta} + (\rho\psi_{1,\rho})_{,\rho} &= \rho
e^{2(\gamma_0-\psi_0)} \left [ {P_0'^2\over R^2}- V(X_0) \right ] \cr
&= {\half}
\rho e^{2(\gamma_0-\psi_0)} ({\cal P}_{_0R} + {\cal P}_{_0\phi}) & (4.15b) \cr
\gamma_{1,\zeta\zeta} + \gamma_{1,\rho\rho} + 2 \psi_{0,\rho} \psi_{1,\rho}
+2 \psi_{0,\zeta} \psi_{1,\zeta} &= - e^{2(\gamma_0-\psi_0)} {\hat T}^\phi_\phi
\cr
&= e^{2(\gamma_0-\psi_0)} {\cal P}_{_0\phi} & (4.15c) \cr}
$$
where ${\cal E}$ and the ${\cal P}$'s are given by (2.10).

We first solve for $\alpha_1$. Note that there is an $\alpha_0=\rho$ in
the $\sqrt{-g}$ on the RHS of (4.15a). This suggests that we 
write
$$
\alpha_1=\rho a(R).
\eqno (4.16)
$$
$a(R)$ then satisfies
$$
a''(R) + {2\over R} a'(R) 
= - [ {\cal E}_{_0} - {\cal P}_{_{0R}} ]
\eqno (4.17)
$$
Thus
$$
\eqalignno{
a(R) &= - \int {1\over R^2} \int R^2 [ {\cal E}_{_0} - {\cal P}_{_{0R}}]dR\cr
&= -\int R [ {\cal E}_{_0} - {\cal P}_{_{0R}}]dR + {1\over R} 
\int R^2 [ {\cal E}_{_0} - {\cal P}_{_{0R}}]dR& (4.18) \cr}
$$
this is readily seen to have the asymptotic form
$$
a(R) \sim -{A\over\epsilon} + {B\over \epsilon R}
\eqno (4.19)
$$
(where $A,B$ are given by (2.16)) and solves the vacuum equations.

Setting $\psi_1=\psi_1(R)$, and using the form of $\alpha_1$ given by
(4.16,18), we see that (4.15b)  becomes
$$
\psi_1''(R) + {1\over R} \psi_1'(R) = {\half} ({\cal P}_{_0R} + {\cal
P}_{_0\phi})  
\eqno (4.20)
$$
which is solved by
$$
\psi_1 = {\half} \int {1\over R} \int R ({\cal P}_{_0R} + {\cal P}_{_0\phi}) 
={\half} \int R{\cal P}_{_0R} 
\eqno (4.21)
$$
using the zeroth order equations of motion (2.12). Thus $\psi_1$ tends
to a constant ($C/2\epsilon$) at infinity which is also a vacuum solution.

Finally, setting $\gamma_1 = \gamma_1(R)$, and using the form of $\psi_1$
given above, (4.15c) reduces to 
$$
\gamma_1''(R) = {\cal P}_{_{0\phi}}
\eqno (4.22)
$$
and
$$
\gamma_1 = \int\hskip -2mm\int {\cal P}_{_{0\phi}} dR 
= \int R{\cal P}_{_{0R}}dR = 2\psi_1.
\eqno (4.23)
$$
Thus, the corrections to the metric written in this form are almost 
identical to the self-gravitating vortex solution. In fact, using these
corrections, we see that the asymptotic form of the
metric given by 
$$
ds^2 \to e^C [e^{2\psi_0} dt^2 - e^{2(\gamma_0-\psi_0)}(dr_c^2 + dz^2) ]
- r_c^2 (1-A+{B\over \sqrt{\lambda}\eta r_c e^{-\psi_0}})^2 
e^{-C} e^{-2\psi_0} d\phi^2 
\eqno (4.24)
$$
in the Weyl metric, or
$$
e^C [ \left ( 1-{2GM\over r_s} \right ) dt^2 - 
\left ( 1-{2GM\over r_s}\right )^{-1} dr_s^2 - r_{_s}^2 d\theta^2]
- r_s^2 (1-A+{B\over \sqrt{\lambda}\eta r_s \sin\theta})^2 e^{-C} 
\sin^2 \theta d\phi^2 
\eqno (4.25)
$$
in the Schwarzschild metric. Note that although the B-term appears to 
distort the event horizon, $B/\sqrt{\lambda}\eta r_s=O(G\mu)\times O(E^{-1})$
and hence represents an effect outside the r\'egime of applicability of
our approximation. We therefore drop this term, rescale the 
metric so that time
asymptotically approaches proper time at infinity, ${\hat t} = e^{C/2}t$
etc.~and setting ${\hat M} = e^{C/2}M$ we have 
$$
ds^2 = \left ( 1-{2G{\hat M}\over {\hat r}_s} \right )
d{\hat t}^2 -  \left ( 1-{2G{\hat M}\over {\hat r}_s}\right
)^{-1} d{\hat r}_s^2 - {\hat r}_s^2  d\theta^2 - {\hat
r}_s^2 (1-A)^2 e^{-2C} \sin^2 \theta d\phi^2 \eqno (4.26)
$$
We thus see that our spacetime is asymptotically locally flat with deficit
angle $2\pi(A+C) = 8\pi G\mu$. Thus by using a physical vortex model, we have
confirmed the results of AFV. However, note that the presence of the radial
pressure term $e^{-C}$ has modified the Schwarzschild mass parameter at infinity
to ${\hat M} = e^{C/2}M$. The gravitational mass of the black hole has
therefore shifted to 
$$
M_g = {\hat M} = e^{C/2}M
\eqno (4.27)
$$
The inertial mass of the black hole, or its internal energy, can be found by
considering the black hole as being formed by a spherical shell of matter
infalling from infinity. Due to the deficit angle this has mass
$$
M_I = {\hat M}(1-A)e^{-C} = M_g(1-4G\mu)
\eqno (4.28)
$$
thus the inertial mass of the black hole is actually less that its
gravitational mass, however, since we cannot accelerate the black hole without
accelerating the string, it is perhaps more correct to refer to this as
the internal
energy of the black hole. We conclude this section on the gravitating
string-black hole system by remarking on the thermodynamics of the system.

Either by euclideanization, or by considering the wave function of a quantum
field propagating on the black hole background, one can see that the
temperature of the black hole is
$T={\tilde \beta}^{-1}=1/8\pi GM_g$. We denote the thermodynamic quantity
$T^{-1}$ as ${\tilde \beta}$ to distinguish it from the Bogomolnyi
parameter, additionally, we have set the Boltzmann constant, k, to unity.
Since the spacetime is
no longer asymptotically flat, euclidean arguments must
be interpreted with care; nonetheless, by a somewhat
non-rigorous partition function calculation we confirm the
AFV result that the entropy of the string-black hole system
is  $$ S = {{\tilde \beta}^2\over 16\pi G}(1-A)e^{-C} =
{1\over 4G} A \eqno (4.29)
$$
Thus although the temperature of the black holes is unchanged in terms of the
gravitational mass measured at infinity, and although the area-entropy
relationship is unchanged, since the internal and gravitational masses are no
longer equal, the entropy of the black hole with the string is less than that of
a black hole of the same temperature (i.e.~gravitational mass) without the
string. 

It is interesting to use these thermodynamical results to examine the dynamical
situation of a cosmic string -- black hole merger. If one demands that the
gravitational mass of the black hole is fixed, then the temperature of the
black hole remains unchanged, but its entropy decreases. If one demands
conservation of internal energy, then the temperature decreases and the entropy
increases. Clearly, thermodynamics indicates that conservation of internal
energy is the correct condition to use. It is interesting to note that in this
case the change in gravitational mass is
$$
\delta M_g = M_g - M_I = 2 \times 2GM_g \times \mu 
\eqno (4.30)
$$
i.e.~the change in gravitational mass is equivalent to the length of string
swallowed up by the black hole as seen from infinity times its energy per unit
length. In this sense, the vortex at infinity is direct hair, conveying exact
information as to the last $4GM\mu$ units of matter that the black hole
swallowed. We will take up this theme further in the next section. 

\vskip 2mm

\noindent{\bf 5. Summary and Discussion}

In this paper we have provided evidence, both analytical and numerical,
that U(1) abelian Higgs vortices can pierce a black hole horizon. We
have shown that there is no gravitational obstruction to this solution,
and at this point it is perhaps worthwhile detailing how our solution
avoids the revamped abelian Higgs no-hair theorem[30]. A simple
answer would be that the string system is not spherically symmetric, however,
many of the steps in [30] can be generalized to include more generic 
situations. Indeed, recent interesting results of Ridgway and Weinberg[31],
who show non-spherically symmetric dressing (although not hair) of
black hole event horizons, indicates that spherical symmetry should not
be a prerequisite of no-hair theorems. The main reason our solution
evades such a no-hair `proof' is that a vortex mandates a non-zero spatial
gauge field, which then destroys the inequalities on which the no-hair
results are based. In particular, the fact that the field $P_\mu$ has lines
of singularity (corresponding to the vortex cores) explicitly breaks the 
argument given in [30] as to the vanishing of $P_i$.

But is the vortex dressing or hair? The thermodynamical argument
seems to indicate that it is hair -- telling us about the last
$4GM\mu$ units of mass the black hole swallowed. Can we make this 
argument stronger? Suppose instead of considering a vortex threading
a black hole we consider a single vortex terminating on a 
black hole. This has a gravitational counterpart in the guise of 
a uniformly accelerating black hole connected to infinity by a 
conical singularity[32], therefore we can ask whether there exists 
a particle physics vortex counterpart to this setup. One immediate 
difference with the previous situation is that the metric here is 
non-static, however, that can be remedied by introducing a second 
black hole attached to infinity by a second string placed so that its
gravitational attraction neutralizes the uniform acceleration[33]. 
This leaves us with the topological question of how to paint a single 
semi-infinite vortex onto a black hole event horizon. Recall in 
section two, when the transformation
to the real variables $X$ and $P_\mu$ was performed that the phase of the Higgs
field, $\chi$, was purely gauge, and only acquired physical significance via
boundary conditions on $P_\mu$. It was remarked that if spatial topology was
trivial, any surface spanning a loop with a winding of $\chi$ would have a
vortex on it. If the space is non-trivial, then the question of whether any
surface spanning such a loop must have a vortex reduces to a question of
topology: if the first Chern class of the U(1) bundle is trivial, then any
spanning surface must have a vortex. 

Recall that in Kruskal coordinates, the extended Schwarzschild spacetime
contains a wormhole: the $t=$constant surface. This has topology
S$^2\times$\real \  with two asymptotically flat regions. Thus the spatial
topology of the Schwarzschild black hole {\it is} non-trivial, and 
is homotopically
equivalent to a sphere. The issue of whether a vortex can terminate on a black
hole therefore reduces to that of placing vortices on 2-spheres -- a well
studied problem (see [34] and references therein). In our case, the 
answer is to take two gauge patches, e.g.
$$\eqalign{
I_1 &= \{\theta,\phi | \theta < \pi/2+\delta\} \cr
I_2 &= \{\theta,\phi | \theta > \pi/2-\delta\} \cr
}
$$
for any $\delta<\pi/2$. Then define the (gauge) transition function
$$
g_{12} = e^{i\phi}
$$
such that
$$
\Phi_1 = g_{12}\Phi_2 \;\;\;\;\; ; \;\;\;\;\;\;\;\;
A_{1\mu} = A_{2\mu} - {g_{12}^{-1}\over e} \partial _\mu g_{12}
$$
on the overlap, and we take $\Phi_2=A_2 = 0$. Thus we have a vacuum on the
southern hemisphere and a vortex on the northern hemisphere connected via a
non-singular gauge transformation on the overlap. Of course, if this two sphere
could be shrunk to zero radius, then this would not be an allowed gauge
transformation, but since the two-spheres in the Schwarzschild spacetime have a
minimum radius, $2GM$, there is no topological obstruction to this definition,
and we can therefore have just a single vortex connected to the black hole. In
terms of the extended Schwarzschild spacetime, this vortex enters the black
hole via the North Pole, goes down the wormhole, and emerges from the North
Pole of the black hole in the other asymptotic regime. The string worldsheet
itself looks like a two-dimensional black hole, but occupies only the
$\theta=0$ portion of the full four-dimensional Penrose diagram. We have not
verified that the Nielsen-Olesen solution can be painted on to the non-static
accelerating black hole spacetime, however, based on the static evidence and
the lack of a topological obstruction,  it would be very surprising if it could
not be. 

This now leaves us with the question of how a black hole might have got just a
single semi-infinite vortex in the first place. Certainly it cannot happen as
the result of interaction between an infinite vortex and a black hole, so let
us consider what the presence of the vortex actually means. When a single vortex
is present on the two-sphere, more than one gauge patch is necessary for a
non-singular description of the physics. This is analogous to the Wu-Yang[35]
description of a Dirac monopole. Indeed, given that in each case we are dealing
with the same mathematical object (a U(1) bundle over the sphere) the only real
difference between the two cases is the spontaneously broken symmetry. Thus the
interpretation of the vortex is that it is localized `magnetic' flux 
emanating from the black hole. In terms of the dynamics of phase transitions in 
the early universe, one is led to a picture of a
magnetically charged Reissner-Nordstrom black hole prior to
the phase transition having its flux localized in the
vortex after the phase transition. Thus the information
(namely `magnetic' charge) which one would not normally
expect to be able to measure corresponding as it does to a
massive field, is indeed preserved for external observers
to see in the form of the long vortex hair stretching to
infinity. We can correspondingly imagine a charge 2
Reissner-Nordstrom hole becoming a Schwarzschild hole with
two vortices extending to infinity from its opposite poles, which would
then be of the AFV form described in section four, where it was the 
energy-momentum rather than the orientation of the vortices that was
relevant. 

In the light of this evidence, we claim that the Abelian Higgs vortex is not
simply dressing of the black hole, as the SU(2) monopole is, but is 
true hair - carrying information from the
black hole to infinity.

\bigskip {\it Note added in proof.} 
After this work was completed, we were informed that Eardley {\it et.al.}[36]
had also developed the gauge patch description (Sec.\ V) in order to argue the
instability of NO vortices to black hole nucleation. Additionally, the
conjecture in Sec.\ V that the NO vortex could be painted on to
the nonstatic C-metric has since been verified in [37].

\bigskip\noindent{\bf Acknowledgements.} 

We have benefited from discussions with many colleagues, in particular Fay
Dowker, I\~nigo Egusquiza, Gary Horowitz, Patricio Letelier,
Nick Manton and Bernd Schroers. 
We wish to thank
the Isaac Newton Institute and the University of Utrecht
for their hospitality. This work was partially supported by
the Isaac Newton Institute, by NSF grant PHY-9309364, CICYT
grant AEN-93-1435 and the University of the Basque Country
grant UPV-EHU 063.310-EB119/92.

\vskip 2mm
\noindent{\bf References.}

\nref
P.T.Chru\'sciel, \book No hair theorems - Folklore, Conjectures, Results [[
gr-qc 9402032]]

\nref 
R.Bartnick and J.McKinnon, \prl 61 41 1988.

\nref
M.S.Volkov and D.V.Gal'tsov, \sovj 51 1171 1990.

\nref 
P.Bizon, \prl 64 2844 1990.

\nref
H.P.K\"unzle and A.K.M.Masood-ul-Alam, \jmp 31 928 1990.

\nref
N.Straumann and Z.-H.Zhou, \plb 237 353 1990. \plb 243 33 1990.

\nref
P.Bizon and R.M.Wald, \plb 267 173 1991.

\nref
M.Bowick, S.Giddings, J.Harvey, G.Horowitz and A.Strominger, \prl 61 2823 1988.

\nref
L.Krauss and F.Wilczek, \prl 62 1221 1989.

\nref
S.Coleman, J.Preskill and F.Wilczek, \mpla 6 1631 1991.
\prl 67 1975 1991.

\nref 
H.F.Dowker, R.Gregory and J.Traschen, \prd 45 2762 1992.

\nref
Adler and Pearson \prd 18 2798 1978.

\nref
G.Gibbons, \book Selfgravitating monopoles, global monopoles and black holes [[ Lisbon Autumn School 1990, p.110]]

\nref
K.Lee, V.P.Nair and E.J.Weinberg, \prd 45 2751 1992. \prl 68 1100 1992.

\nref
M.Aryal, L.Ford and A.Vilenkin, \prd 34 2263 1986.

\nref
H.B.Nielsen and P.Olesen, \npb 61 45 1973.

\nref
E.B.Bogomolnyi, {\it Yad.~Fiz.} {\bf24} 861 (1976)[{\it
Sov.~J.~Nucl.~Phys.~{\bf 24}} 449 (1976)]

\nref
K.S.Thorne, \prb 138 251 1965.

\nref
R.Gregory, \prl 59 740 1987.

\nref
D.Garfinkle, \prd 32 1323 1985.

\nref
A.Vilenkin, \prd 23 852 1981.

\nref
J.R.Gott III, \apj 288 422 1985.

\nref
W.Hiscock, \prd 31 3288 1985.

\nref
B.Linet, \grg 17 1109 1985.

\nref
J.L.Synge, \book Relativity: The General Theory [[ Amsterdam North Holland,
1960]]

\nref
A.Ach\'ucarro, R.Gregory, J.Harvey and K.Kuijken, \prl 72 3646 1994.

\nref
R.Gregory, \plb 215 663 1988.

\nref
G.Gibbons, M.Ortiz and F.Ruiz, \prd 39 1546 1989.

\nref
W.H.Press, B.R.Flannery, S.A.Teukolsky and W.T.Vetterling, \book 
Numerical Recipes [[Cambridge University Press, 1988]]

\nref
A.Lahiri, \mpla 8 1549 1993.

\nref
S.A.Ridgway and E.J.Weinberg, {\it Static Black Hole Solutions 
Without Rotational Symmetry}, gr-qc/9503035, CU-TP-673.

\nref
W.Kinnersley and M.Walker, \prd 2 1359 1970.

\nref
W.Israel and K.A.Khan, \ncim 33 331 1964.

\nref
N.Manton, \npb 400 624 1993.

\nref
T.T.Wu and C.N.Yang,  \prd 12 3845 1975. \npb 107 365 1976.

\nref
D.Eardley, G.Horowitz, D.Kastor and J.Traschen, \prl 75 3390 1995. 
gr-qc/9506041.

\nref 
R.Gregory and M.Hindmarsh, \prd 52 5598 1995. gr-qc/9506054
\vfill\eject

\input epsf.tex

\noindent Figure 1. Numerical solution of the Nielsen-Olesen equations
with $N=1$, ${\beta}=1/2$ in a Schwarzschild metric
($E=10$) background. The event horizon is indicated by a
semicircle. Evidently the presence of the black hole
horizon hardly affects the string structure at all. This
solution was calculated with 100 radial and 100 azimuthal
grid points, out to radius $r_L=60$.

\vskip -75truemm
\epsfxsize=15cm
\epsfbox{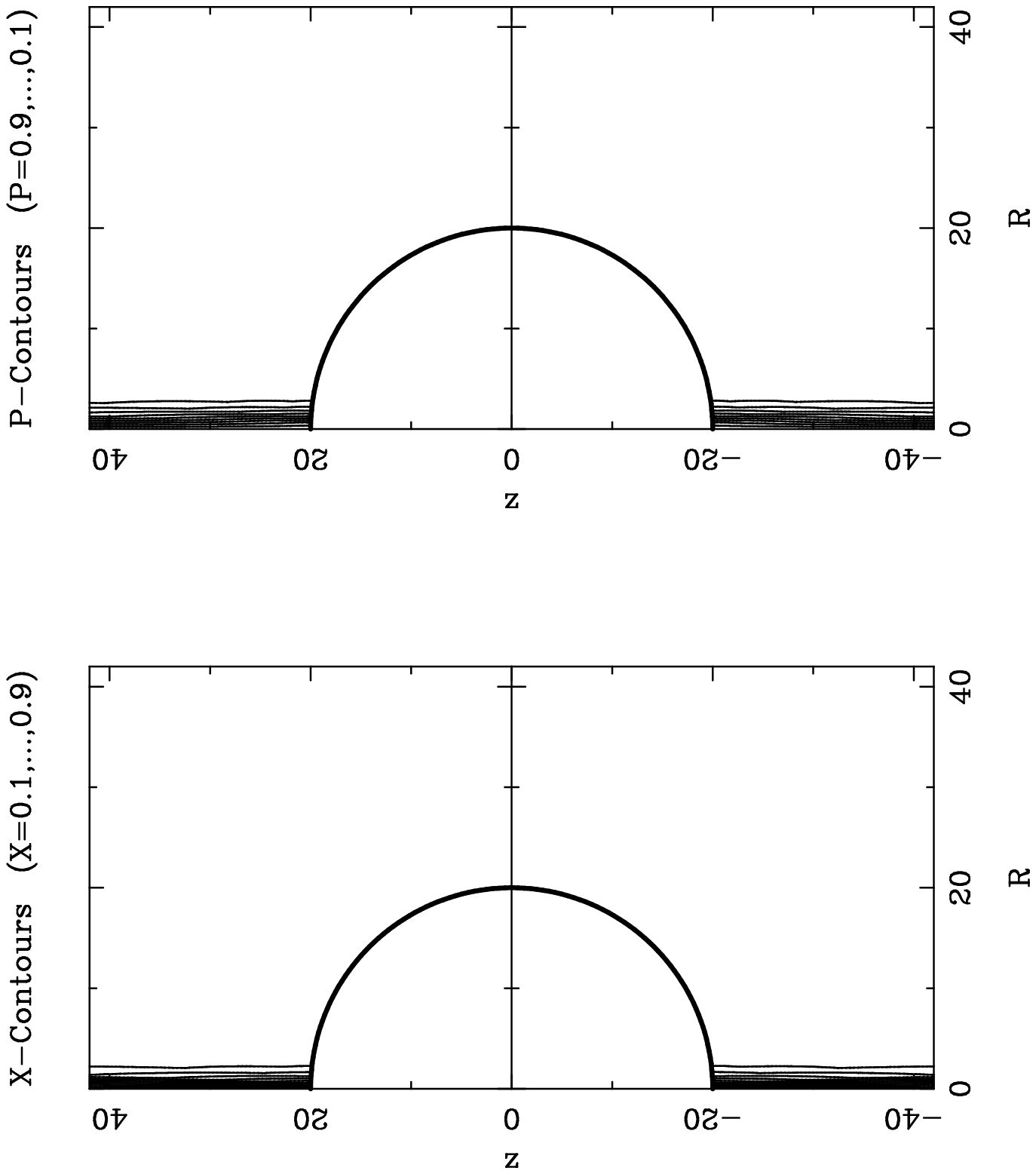}
\vfill\eject

\noindent Figure 2. As figure 1, but for winding number 5. 

\vskip -75truemm
\epsfxsize=15cm
\epsfbox{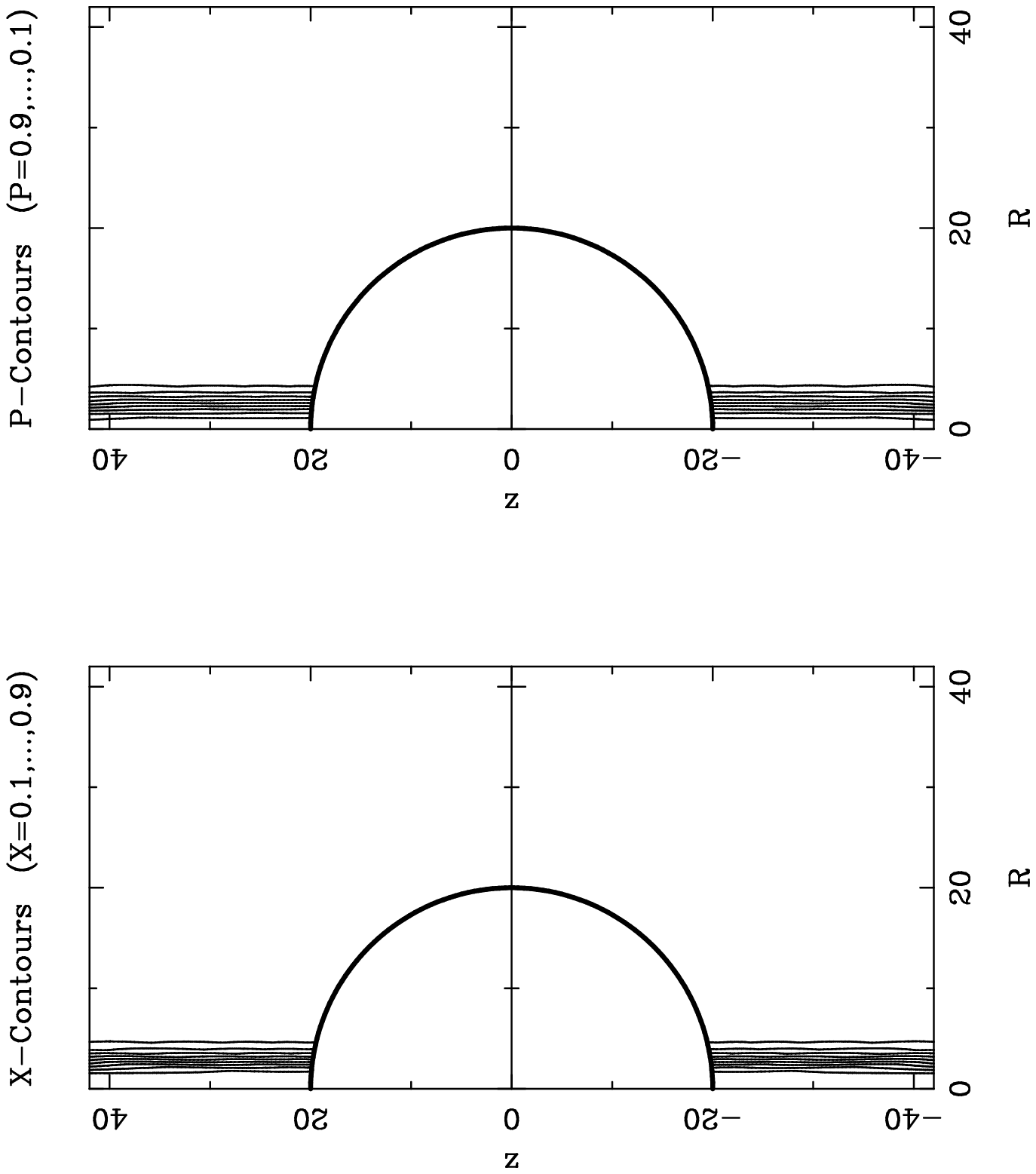}
\vfill\eject

\noindent Figure 3. As figure 1, but for winding number 100. The
string is noticeably pinched. ($r_L=100$ for this calculation.) The
undulations in the outer contours occur on the scale of one grid cell,
and are an artifact of the contouring package's conversion from polar
to rectangular coordinates.

\vskip -75truemm
\epsfxsize=15cm
\epsfbox{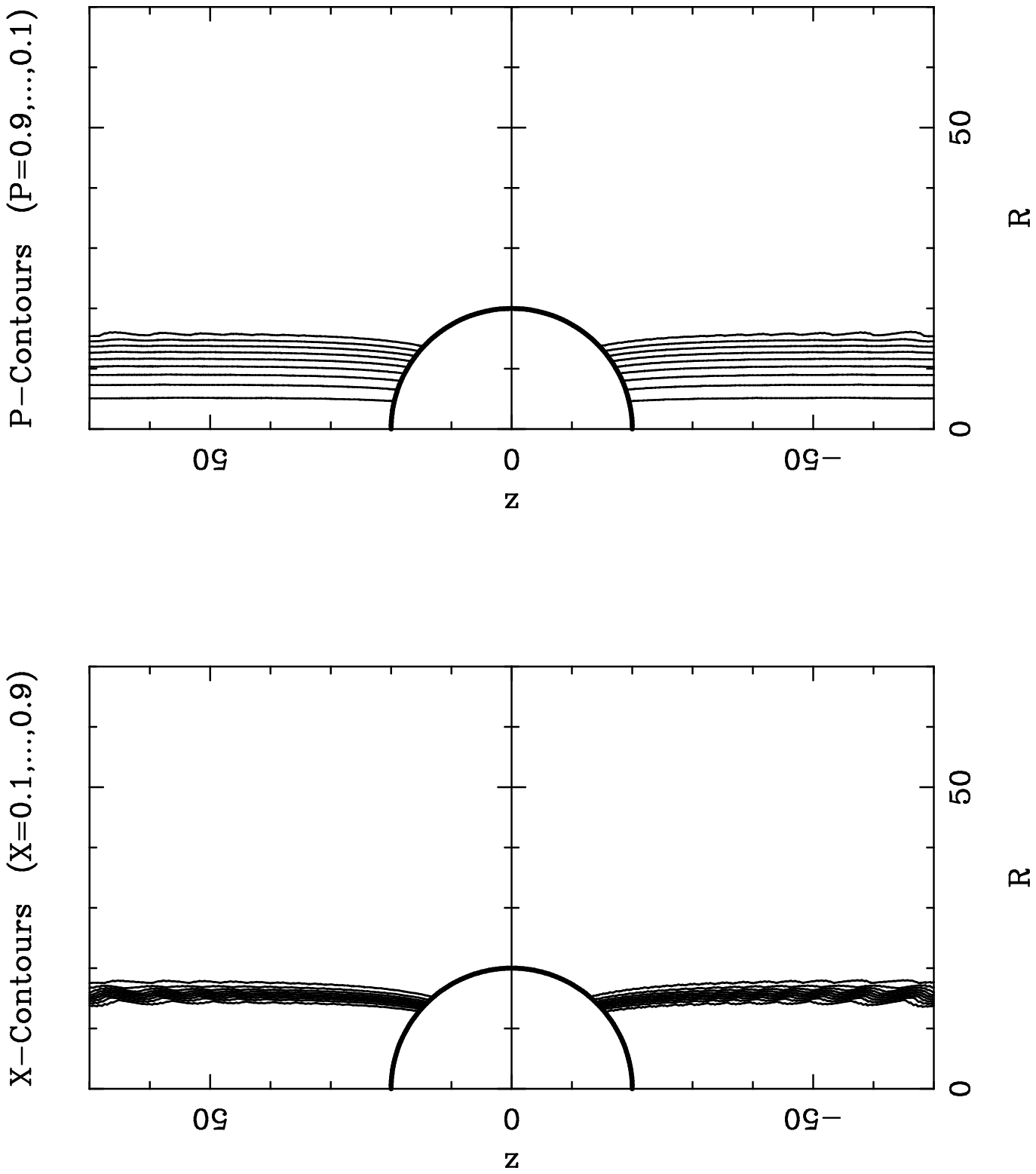}
\vfill\eject

\noindent Figure 4. As figure 1, but for winding number 400. The event
horizon is now entirely inside the core of the string, which is
slightly pinched. ($r_L=150$)

\vskip -75truemm
\epsfxsize=15cm
\epsfbox{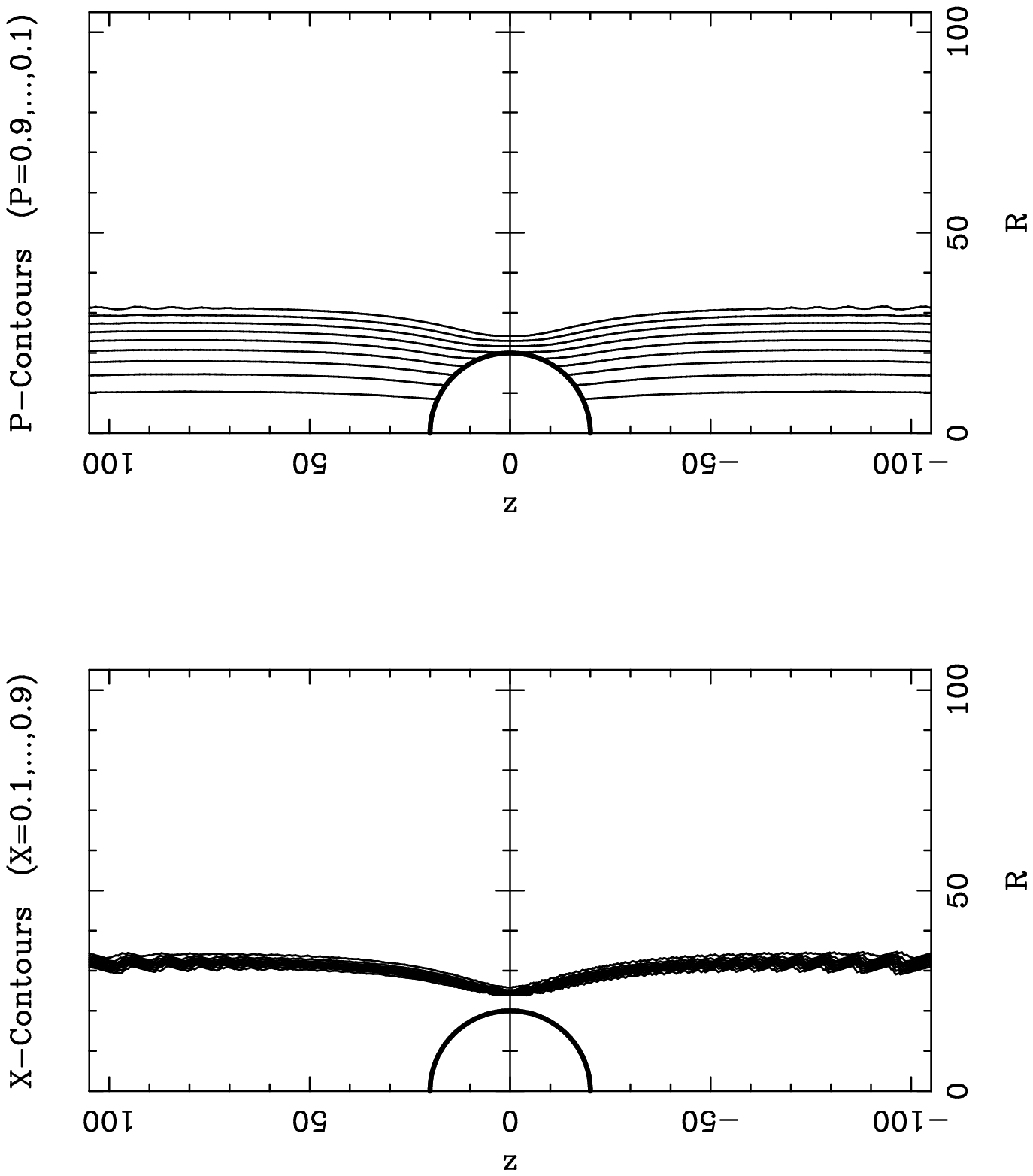}
\vfill\eject

\noindent Figure 5. As figure 1, but with $E=1$, $N=1$, $r_L=15$. String and
black hole have comparable radii, but distortion of the string by the
background is still rather mild.

\vskip -75truemm
\epsfxsize=15cm
\epsfbox{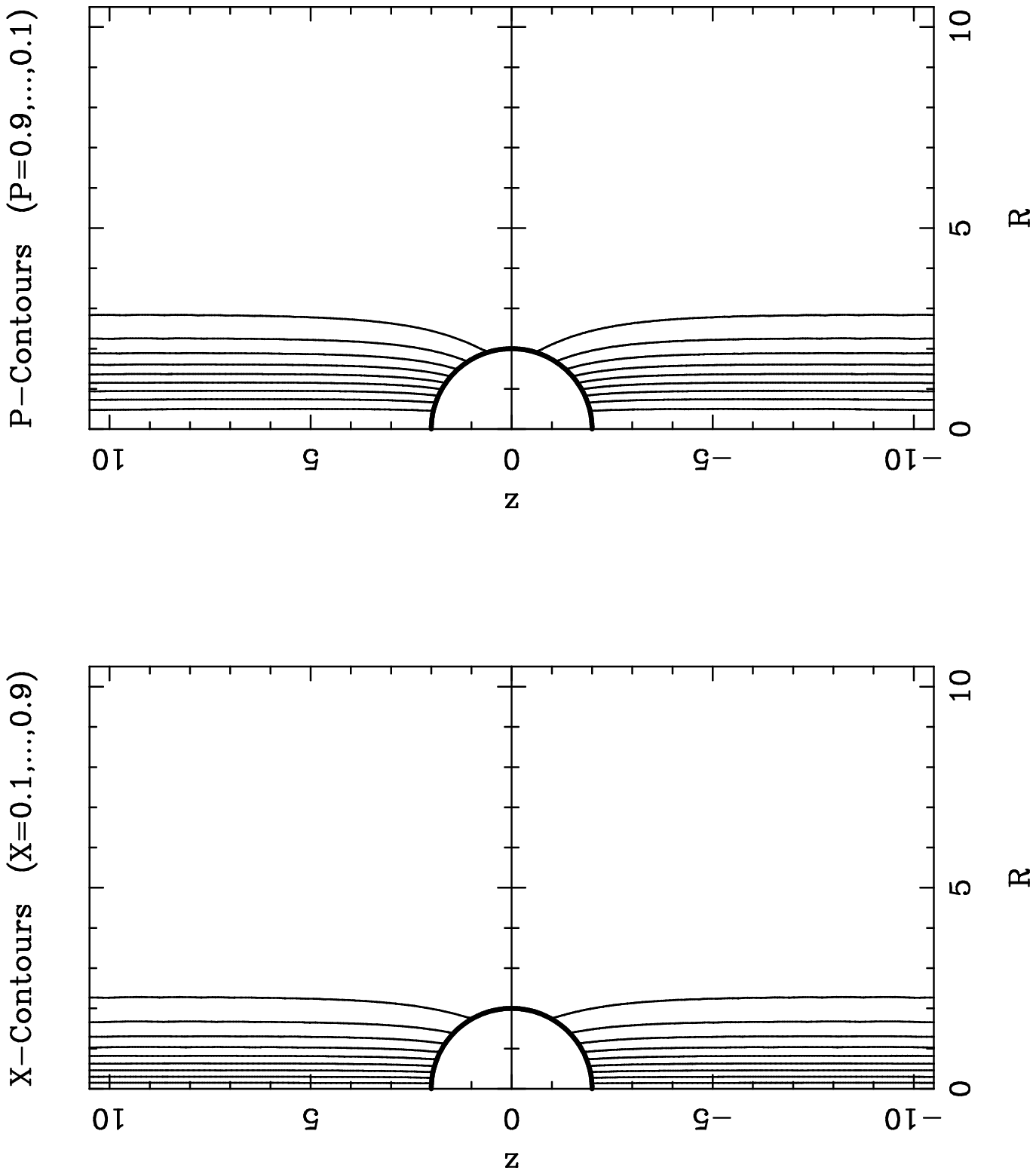}
\vfill\eject

\noindent Figure 6. As figure 5, but for a global string. In agreement with
our analytical arguments, the effect of the black hole on a global string
is weaker than on the corresponding local string.

\vskip -75truemm
\epsfxsize=15cm
\epsfbox{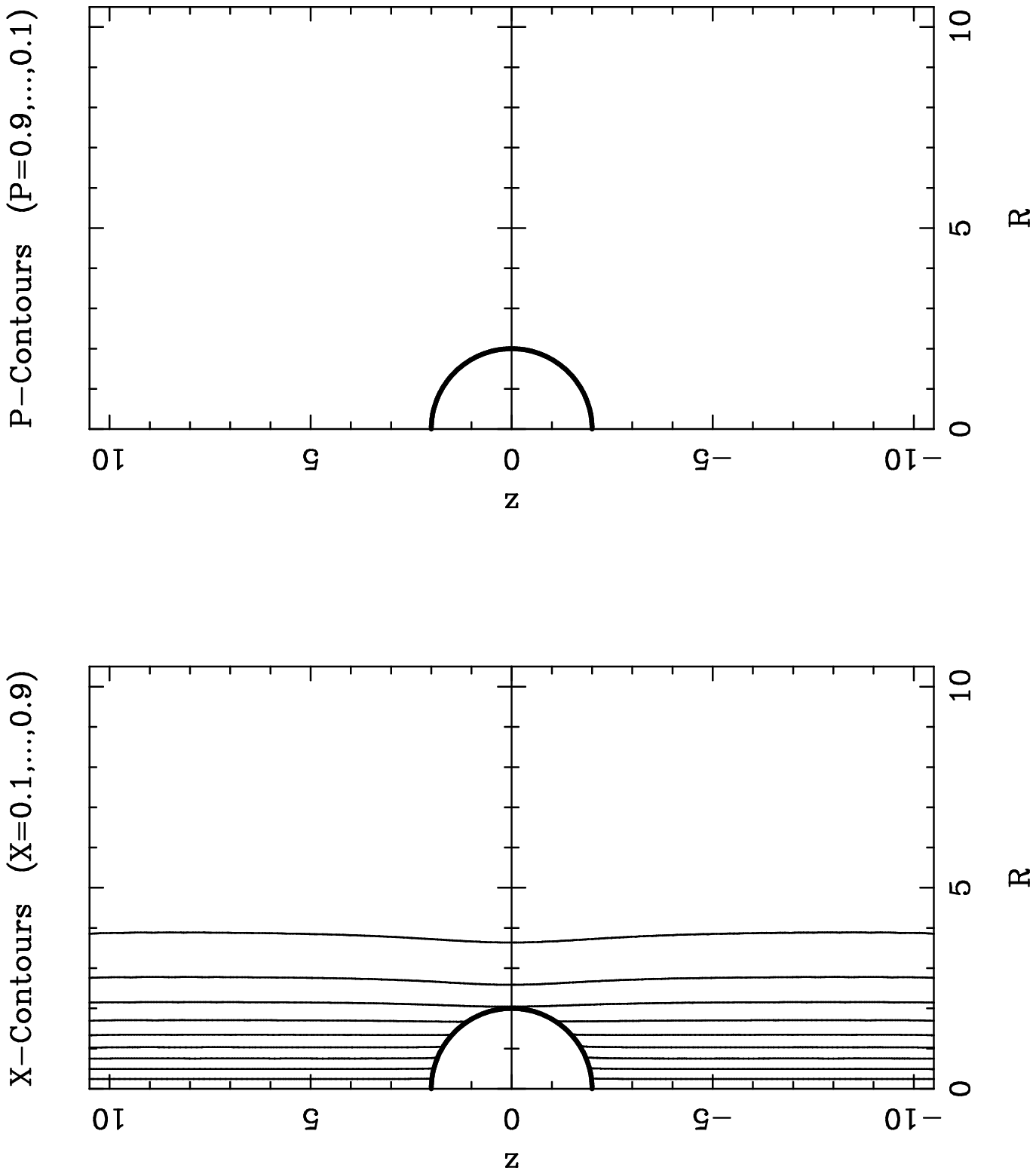}
\vfill\eject

\epsfxsize=15cm
\epsfbox{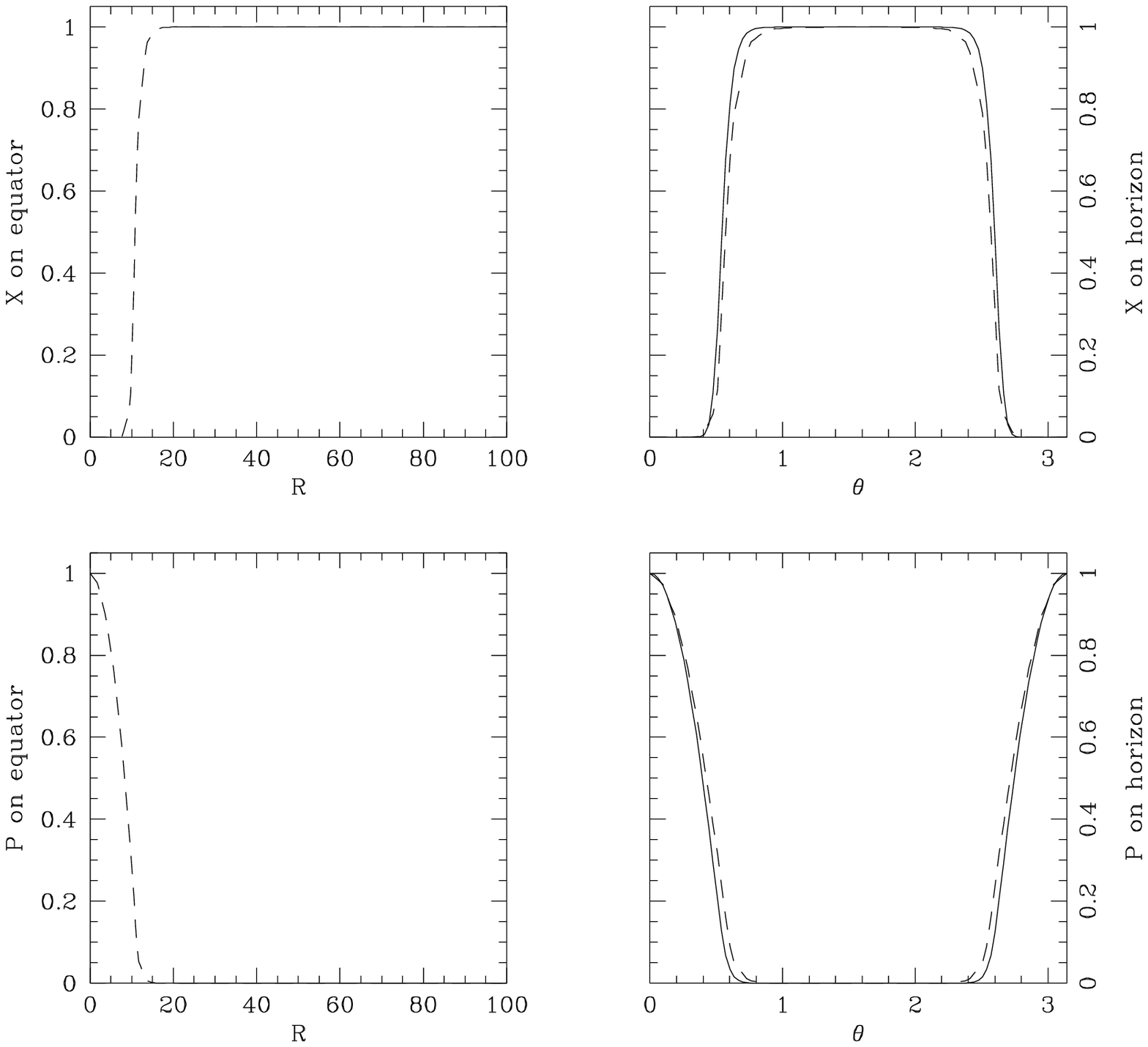}
\noindent Figure 7. Illustration of the relatively small effect of a black
hole horizon ($E=10$) on a local ($\beta=0.5$, $N=50$) string. In these
panels, solid lines show values of the field in the black hole
background, and dashed lines the 
values at corresponding positions in a flat metric. Upper and lower
panels show $X$ and $P$, respectively, while left and right panels
show cuts along the equator ($\theta=\pi/2$) and around the horizon
($r=2E$).

 \bye